# Current-driven magnetization dynamics and their correlation with magnetization configurations in perpendicularly magnetized tunnel junctions

Kaiyuan Zhou[1], Lina Chen[2], Kequn Chi[3], Qingwei Fu[1], Zui Tao[1], Like Liang[1], Zhenyu Gao[1], Haotian Li[1], Hao Meng[3], Bo Liu[3], Tiejun Zhou[4], and R.H. Liu[1,*]

[1]National Laboratory of Solid State Microstructures, School of Physics and Collaborative Innovation Center of Advanced Microstructures, Nanjing University, Nanjing 210093, China
[2]School of Science, Nanjing University of Posts and Telecommunications, Nanjing 210023, China.
[3]Key Laboratory of Spintronics Materials, Devices and Systems of Zhejiang Province, Zhejiang 311305 China.
[4]Centre for Integrated spintronic devices, School of Electronics and Information, Hangzhou Dianzi University, Hangzhou 310018, China

We study spin-transfer-torque driven magnetization dynamics of a perpendicular magnetic tunnel junction (MTJ) nanopillar. Based on the combination of spin-torque ferromagnetic resonance and microwave spectroscopy techniques, we demonstrate that the free layer (FL) and the weak pinned reference layer (RL) exhibit distinct dynamic behaviors with opposite frequency vs. field dispersion relations. The FL can support a single coherent spin-wave (SW) mode for both parallel and antiparallel configurations, while the RL exhibits spin-wave excitation only for the antiparallel state. These two SW modes corresponding to the FL and RL coexist at an antiparallel state and exhibit a crossover phenomenon of oscillation frequency with increasing the external magnetic field, which could be helpful in the mutual synchronization of auto-oscillations for SW-based neuromorphic computing.

* Corresponding author, Email: rhliu@nju.edu.cn

## I. INTRODUCTION

The dc current can control the magnetization state or drive the stable procession of magnetization in a sandwiched magnetic structure by the spin-transfer torque (STT) effect because the unpolarized electrons passing through the two magnetic layers separated by a nonmagnetic metal or ultra-thin insulating layer first are polarized by the fixed magnetic layer and then transfer the spin angular momentum to the free layer[1,2]. The STT is the critical factor in the current-driven magnetization reversal and auto-oscillation by compensating for the intrinsic damping and developing the so-called magnetic random access memory (MRAM) and nano-oscillator[3-6]. Compared to traditional semiconductor memory, STT-MRAM is a next-generation non-volatile memory technology with many advantages of low power consumption, high speed, and almost unlimited endurance[7]. Meanwhile, the STT-induced magnetization precession enables the MRAM cell to be a tunable microwave nano-oscillation, named spin-transfer nano-oscillator (STNO). STNOs exhibit rich nonlinear magnetic dynamics and have potential applications in wireless communications and neuromorphic computing because their amplitude, frequency, and phase can be tuned by applied current and magnetic field[8-15].

Modern STT-MRAM is based upon nanoscale MgO-CoFeB-based MTJs with high tunneling magnetoresistance (TMR) and strong perpendicular magnetic anisotropy (PMA). Compared to in-plane MTJs, perpendicular MTJs (pMTJs) have a relatively low critical current and high thermal stabilization, providing a route toward scalable STT-MRAM technology[16-21]. Besides the advantages of STT-MRAM, pMTJ-based STNO is also a promising candidate for next-generation ratio-frequency (rf) technology because it is over 50 times smaller than a standard voltage-controlled oscillator based on complementary metal-oxide-semiconductor (CMOS) technology[22], low biased voltage, wide working temperature, and tunable frequency range over dozens of GHz. With the development of 5G wireless communications, there are increasing need for this low power consumption, high speed, and multi-function next-generation rf technology[23-25]. Therefore, it is essential to fully understand the magnetic dynamics and corresponding microwave emission characteristics of MgO-CoFeB-based STNO with a high TMR.

Here, we report an experimental study of STT-driven magnetization dynamics in pMTJ nanopillar with sufficient PMA and as high as TMR ~180% at a low bias voltage based on CoFeB/Ta/CoFeB as the free layer (FL), synthetic ferrimagnet (SyFi) multilayer [Pt/Co]/Ru/[Pt/Co] as the pinning layer (PL), and CoFeB as the reference layer (RL). The field-driven magnetization switching measurements show that the studied MTJ cells exhibit a relatively moderate interlayer exchange coupling (IEC) between SyFi-PL and RL through a relatively thick Ta spacer, causing a positive bias magnetic field $H_{bias}$ on the adjacent CoFeB-RL. Furthermore, by combining spin-torque ferromagnetic resonance (ST-FMR) and microwave spectroscopy techniques, we find that this pMTJ exhibits the two distinct dynamic modes with the opposite frequency vs. field dispersion relations, corresponding to the dynamics of the FL and the RL layers, respectively.

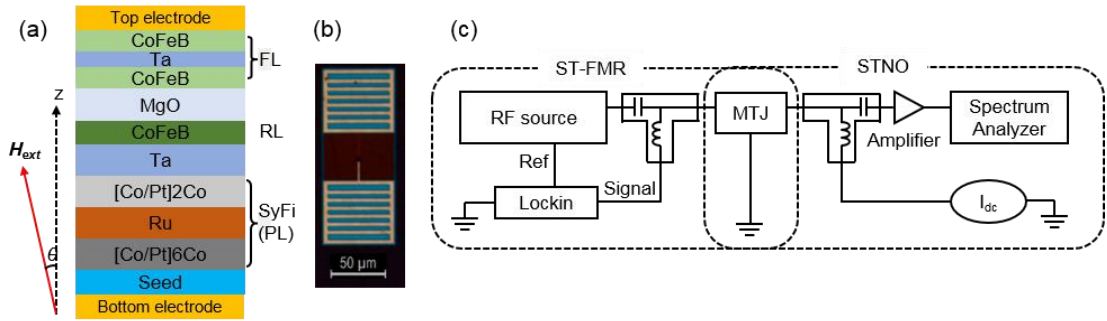

FIG.1. (a).The schematic of the multilayer stack structure of MTJ. (b). The optical images of the single MTJ cell and its two external electrodes. (c). Schematic illustration of the device and the experimental setup of ST-FMR and microwave spectroscopy. The external magnetic field $H_{ext}$ is applied with an angle $\theta$ relative to the normal direction (z-axis) of the MTJ multilayer stack, as shown in the left inset of (a).

## II. EXPERIMENTAL TECHNIQUES

The pMTJ nano-pillar with 70 nm diameter were nanofabricated by combining photolithography and ion beam milling technique based on the multilayer stack of Ta(5) / Pt(5) / [Co(0.4)/Pt(0.4)]$_6$ / Co(0.4) / Ru(0.9) / [Co(0.4)/Pt(0.4)]$_2$ / Co(0.4) / Ta(0.55) / Co$_{20}$Fe$_{60}$B$_{20}$(1) / MgO(0.9) / Co$_{20}$Fe$_{60}$B$_{20}$(1.6) / Ta(0.43) / Co$_{20}$Fe$_{60}$B$_{20}$ (1.0) / MgO(0.9)

/ Ta(5) / Ru(5), as shown in Fig. 1(a). The number in parenthesis is the thickness in nm. The size of two external electrodes of the studied pMTJ cell is shown in Fig. 1(b). The [Co/Pt] multilayers are coupled in antiferromagnetic by 0.9 nm thick Ru spacer and form the SyFi pinning layer. The top adjacent CoFeB(1) as the RL couples with the bottom SyFi-PL through 0.55 nm Ta spacer. Figure 1(c) illustrates the experiment setup of ST-FMR and generation microwave spectroscopy. A Lock-in amplifier detects the dc voltage of ST-FMR spectra with a reference signal (2 kHz) to modulate the amplitude of the output *rf* signal from a signal generator. A spectrum analyzer with a 30 dB external gain amplifier is used to detect the generation microwave signal of the device for the analysis of its dc current-driven magnetodynamics.

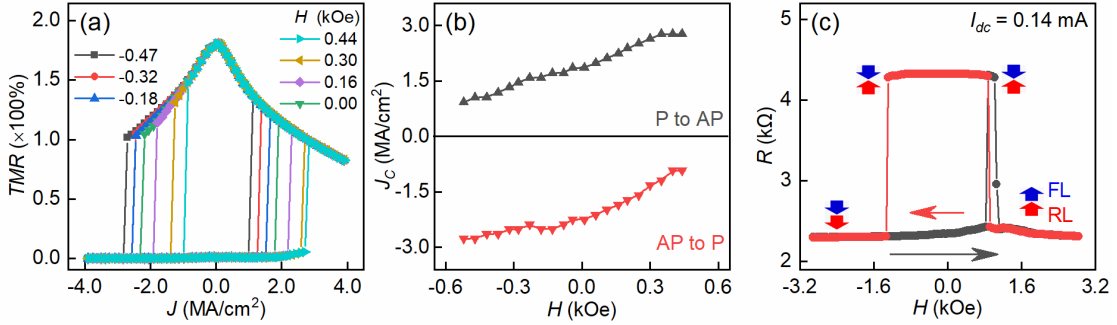

FIG.2. (a). TMR hysteresis loops as a function of dc current density $J$ at different labeled out-of-plane magnetic fields ($\theta = 0°$). (b). The critical switching current density $J_c$ vs. $H$ for P to AP (top panel) and the AP to P reversals (bottom panel). (c). The major TMR hysteresis loop of MTJ with $I_{dc} = 0.14$ mA. The black and red arrows indicate the sweeping field directions, and the bold blue and red arrows represent the FL and RL magnetization directions, respectively.

## III. RESULTS AND DISCUSSION

Firstly, the current-driven magnetization switching experiments are adopted to investigate STT switching behavior in our MgO-CoFeB-based pMTJs. Figure 2(a) shows the representative TMR vs. the current density J loops at different out-of-plane magnetic fields varied from −0.47 to 0.44 kOe. The TMR ratio is defined as TMR=100%×($R_{AP}$-$R_P$)/$R_P$, where $R_P$ and $R_{AP}$ are the resistances in the parallel (P) and antiparallel (AP) magnetization alignments of RL and FL, respectively. The TMR ratio shows a noticeable decline with increasing the bias current due to spin excitation localized at the interfaces between the CoFeB magnetic layers and the MgO tunnel barrier, consistent with the previous reports [21, 26-29]. Figure 2(b) shows that the critical current density $J_c^{AP \to P}$ (from AP to P state) is larger (less) than $J_c^{P \to AP}$ (from P to AP state) when applied a negative (positive) perpendicular magnetic field. These field-dependent critical current density behaviors are consistent with the previous studies[30-32]. The average $J_C = (J_c^{AP \to P} - J_c^{P \to AP})/2 = 3.0 \text{ MA}/cm^2$ is comparable to previously reported CoFeB/MgO-based MTJs[33-35]. $J_C$ can be expressed quantitatively as the follows:

$$J_C \approx \frac{2e}{\hbar} \cdot \frac{t_F \alpha M_S}{\eta} \cdot \left(H_{ext} - 4\pi M_{eff}\right) \quad (1)$$

where $e$ is the electron charge, $\hbar$ is the reduced Planck constant, $\alpha, M_S$ and $t_F$ are the damping factor, the saturation magnetization and thickness of the FL, $\eta$ is the effective spin-transfer efficiency. The effective magnetization $4\pi M_{eff}$ can be further rewritten as $4\pi M_s - \frac{2K_\perp}{M_s}$. According to Eq.1, the fitting slope of the experimentally obtained $J_C$ vs. $H_{ext}$ data is $0.27 \, MA/(cm^2 \cdot kOe)$. This slope value can be used to accurately determine the effective spin-transfer efficiency $\eta = 0.3$ after extracting the damping factor $\alpha$ of the FL from the frequency-dependence of the FMR linewidth obtained by our broadband ST-FMR technique below.

Furthermore, Figure 2(c) shows the representative major TMR hysteresis loop under out-of-plane magnetic fields with $I_{dc}$ = 0.14 mA. The sharp switching in the TMR loop suggests that the studied pMTJ has a well-defined strong PMA for both RL and FL. Based on our previously reported analysis method of the field-dependent asymmetric major and minor loops[36], we can quantitatively determine the offset field $H_{offset} = -0.96 \, kOe$ due to the stray field of the bottom SyFi layer and the bias field $H_{bias} = 1.16$ kOe on the RL due to the IEC between the RL and SyFi layer in this device. The stray field can also theoretically be estimated at -0.86 kOe using COMSOL Multiphysics, consistent with -0.96 obtained from R vs. H loops.

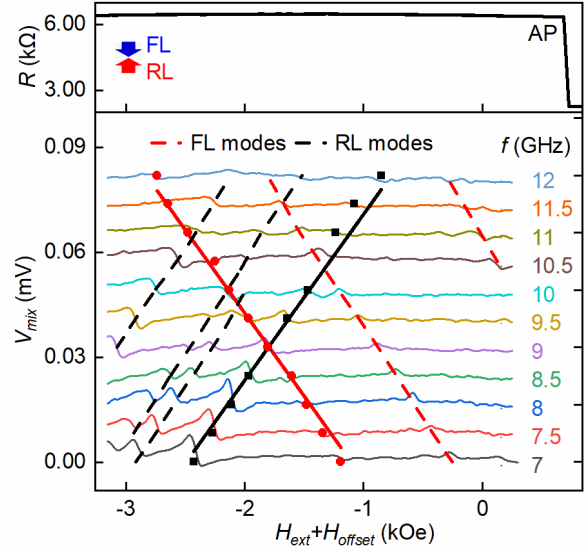

FIG.3. Top panel: the resistance of the studied pMTJ as a function of the applied perpendicular magnetic field at $I_{dc} = 1\ \mu A$. The bold blue and red arrows illustrate the magnetization of the FL and RL, respectively. Bottom panel: the ST-FMR spectra with different excitation frequencies from 7 to 12 GHz in 0.5 GHz steps at the AP state. The solid lines are the fitting of the primary uniform FMR modes for the FL(red circle) and RL(black square). The dashed lines are given as guides to the eye for the dispersion relations of other modes.

Now, we use the ST-FMR technique to characterize the magnetic dynamics of pMTJ nanopillar[37]. The ST-FMR measurements are performed by scanning $H_{ext}$ perpendicular to the device plane at different excitation frequencies. To better allow us to do quantitative analysis, ST-FMR spectra are only conducted at the AP state during $H_{ext}$ range from -2.5 kOe to 1.25 kOe, which was confirmed by the R vs. H curve in the top panel of Fig.3. The ST-FMR spectra can be fitted by the sum of symmetric and anti-symmetric Lorentzians[38]

$$V_{mix} = S \frac{\Delta H^2}{\Delta H^2 + (H_{ext} - H_r)^2} + A \frac{\Delta H(H_{ext} - H_r)}{\Delta H^2 + (H_{ext} - H_r)^2} \quad (2)$$

where $\Delta H$ is the linewidths, $H_r$ is the resonance field, S and A are the symmetric and antisymmetric components of Lorentzian function, respectively. The several dispersion curves of f vs. H are observed in ST-FMR spectra under the AP state. Three of them have the resonance frequency $f$ decreasing with $H_{ext}$ increasing, indicating that the corresponding total effective magnetization decreases with increasing $H_{ext}$. These results are consistent with the fact that the FL (RL) magnetization direction is antiparallel (parallel) to the total effective field, which is confirmed by $R$ vs. $H_{ext} + H_{offset}$ curve in the top panel of Fig.3. Therefore, we can identify those modes with a negative (positive) dispersion as the dynamical modes of the FL (RL) layer[39]. Contrary to the absent FMR mode in previously reported pMTJs with the well-pinned RL[20, 36], the observed substantial FMR modes for the RL suggest that the pMTJ has a relative weak IEC between the RL and SyFi-PL, consistent with our prior field-driven magnetization reversal experiments[36] and previous results in pMTJs without IEC where the external *rf* signal and/or current-induced STT also can drive dynamic mode excitation in the RL[39].

Since the nanopillar shape induced the spatial confinement effect on the dynamic modes, additional localized standing wave modes are also observed. In principle, the frequency dispersions of these modes can also be simply described by Kittel formular[39]:

$$\hbar\omega = 2\mu_B(H_r + H_{stray} + 4\pi M_{eff}) + Dk^2 \quad (3)$$

where $k$ is the spin-wave vector, D is the exchange-stiffness, and $H_{stray}$ is the total effective field on the active dynamic layer, including the stray field and IEC-induced bias field from the SyFi layer. Here, we use the FMR mode of the FL as an example to illustrate our analysis process. Except for the external magnetic field, the FL also suffers $H_{stray}$ layer in the multilayer structure. The sum of the stray field and the effective magnetization $H_{stray} + 4\pi M_{eff} = -2.269\ kOe$ can be determined by fitting the negative experimental dispersion curves with Eq.3 above. Effective magnetization is defined as $4\pi M_{eff} = 4\pi M_s - H_\perp$, where $M_s = 12.6\ kOe$ is the saturation magnetization, and $H_{stray}$ on the FL is -0.96 kOe. Therefore, we can determine the PMA field $H_\perp = 13.88\ kOe$ of the FL. The PMA field $H_\perp = 11.32$ kOe of the RL also can be examined by this similar process based on its FMR dispersion data. Except to determine the total effective magnetic field on the FL and RL individually, the Gilbert damping constant also can be extracted from the frequency-dependent ST-FMR spectra by using $\Delta H = \Delta H_0 + \frac{\alpha}{\gamma}\omega$. The obtained Gilbert damping constant $\alpha$ is 0.013 for the FL and 0.010 for the RL, respectively.

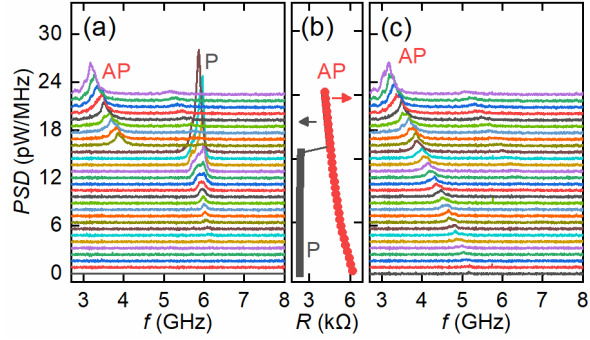

FIG.4. Dependence of the microwave generation characteristics on the excitation current at $H_{ext}$ = 0.44 kOe and $\theta$ = 45°. (a) and (c) Microwave-generation spectra obtained with increasing current $I$ from 0.01 mA (P state) to 0.15 mA (AP state) (a) or decreasing current $I$ from 0.15 mA to 0.01 mA in 5 µA steps (always stay in AP state) (c). (b) The corresponding resistance of the device during spectroscopic measurements.

To explore the dynamical states induced by dc current in this pMTJ, we perform microwave spectroscopic measurements with varying the tilted angle $\theta$ of $H$ relative to the multilayer plane. Tilting $H$ from the vertical direction can enhance the amplitude of the procession and improves the output power of the microwave generation signal. Figure 4 shows representative auto-oscillation spectra acquired at $H$ = 0.44 kOe, $\theta$ = 45° with sweeping $I$ from 0.01 mA to 0.15 mA [Fig.4(a)] and then back to 0.01 mA [Fig.4(c)] in 5 µA steps. From the R vs. $I$ loop in Fig.4(b), the magnetization of FL switches to the high resistance AP state from the low resistance P state at $I_t$ = 0.11 mA while increasing $I$ from 0.01 mA to 0.15 mA process due to current-induced STT. It is consistent with the fact that FL and RL prefer the AP configuration when a large current flows to the FL from the RL due to the STT effect. Figure 4(a) shows that the spectrum shows a small peak at 6 GHz above the critical current $I_c$ = 0.04 mA, which is close to the calculated value of 0.023 mA by Slonczewski's model above[40]. Then the oscillation suddenly jumps to a low-frequency $f$ = 4 GHz with a much small power density from 6 GHz at the threshold current $I_t$ = 0.11 mA, which is related to the FL magnetization reversal from P to AP state observed in Fig. 4(b). It is further confirmed by the absent frequency jump with changing current from 0.15 mA to 0.01 mA in Fig.4(c) because the device has a preferable AP as the initial state and will not happen STT-induced magnetization switching of the FL at the positive currents. The power spectral density (PSD) of the observed spectra can be well approximated by a Lorentzian function, indicating a good coherent dynamic mode rather than thermally-excited precession. Note that the observed magnetization auto-oscillation in the AP state at large positive currents is still driven by current-induced STT, because the magnetizations of the RL and FL are not completely antiparallel to each other under a tilted H ($\theta = 45°$).

The auto-oscillation spectra show much higher PSD at the P state than at the AP state, indicating that the FL has a larger precession angle in the P state, which is also consistent with the fact that the polarized electrons prefer to drive the FL magnetization antiparallel to the RL at the positive current due to STT effect. We also find that the oscillation frequency exhibits a significant red shift with the current in the AP state (15.75 MHz/μA) and a little shift in the P state, indicating that these two states have significantly different nonlinear coefficients.

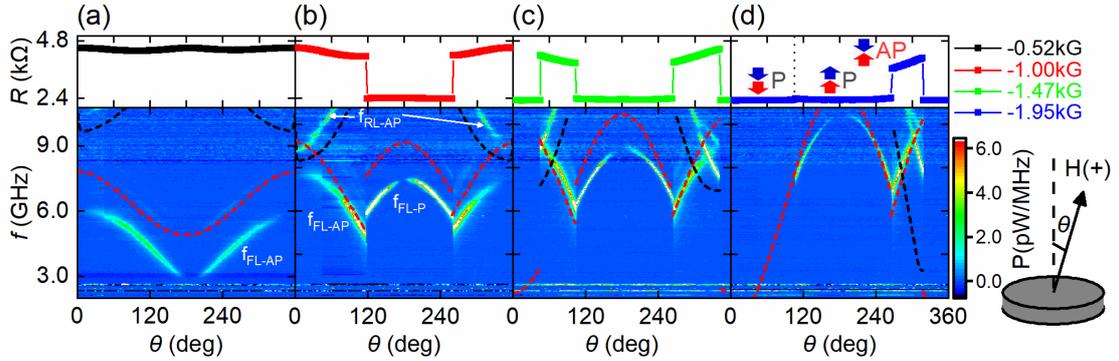

FIG.5. Dependence of the microwave generation characteristics on tilted angle $\theta$ at $I = 0.12$ mA and several fields. (a) – (d). $R$ vs. $\theta$ curves (Top panel), and pseudocolor plots of the spectra (bottom panel) obtained at $H$ = -0.52 kOe (a), -1 kOe (b), -1.47 kOe (c) and -1.95 kOe (d). The bold blue and red arrows in the top inset of (d) illustrate the FL and RL magnetization configurations, respectively. Dashed lines are some experimental results obtained by ST-FMR and theoretically calculated FMR frequency of the FL (red line) and RL (black line) as a function of tilted angle $\theta$ at labeled fields. Right inset: definition of the magnetic field tilted angle $\theta$.

As ST-FMR spectra discussed above, the studied pMTJ device exhibits two distinct types of dynamics corresponding to the FL and RL, respectively. Therefore, to obtain more information on its auto-oscillating dynamics, we further perform dependence of the microwave generation characteristics with a wide frequency range on the magnetic field tilted angle $\theta$ at $I = 0.12$ mA and four different fields $H$ = -0.52 kOe, -1.0 kOe, -1.47 kOe, and -1.95 kOe, as shown in the bottom panel of Fig.5. The top panel of Fig.5 shows the resistance of MTJ as a function of $\theta$ at the four different fields, which can let us to know the magnetization configurations simultaneously. The P and AP states are illustrated in the inset of Fig.5(d).

The obtained microwave generation spectra exhibit two opposite frequency dependences on $\theta$ for all four applied fields, similar to the $f$ vs. $H$ dispersion obtained in the ST-FMR spectra. The low-frequency modes $f_{FL-P}$ and $f_{FL-AP}$ with a larger PSD, corresponding to the oscillating magnetization of the FL, decreases its frequency with increasing $\theta$ in the AP state and reverses the variation tendency after switching to the P state from the AP state above the critical angle (e.g., $\theta = 120°$ for $H = -1.0$ kOe). While the high-frequency mode $f_{RL-AP}$ corresponds to the RL is only observed in the AP state and exhibits an opposite angular dependence of frequency with the primary $f_{FL-AP}$ mode. Since the AP state exhibits the opposite angular dependence for $f_{RL-AP}$ and $f_{FL-AP}$, there exists the oscillation frequency crossover behavior for these two modes when the applied external magnetic field is above 1.47 kOe. From the analysis of angular dependence of the spectra frequency and power, we confirm no synchronization (level attraction) or hybridization gap occurrence at the frequency cross point for these two modes corresponding to the RL and FL, respectively, suggesting a negligible mode coupling between RL and FL in the studied MTJ.

Furthermore, to directly compare the frequency of auto-oscillation and uniform FMR, we theoretically calculate the FMR frequency with the parameters extracted from our experimentally obtained ST-FMR data by using the Kittel model [41]:

$$\left(\frac{\omega}{\gamma}\right)^2 = H_1 H_2$$

$$H_1 = H_r \cos(\theta_H - \theta_M) - 4\pi M_{eff} \cos(2\theta_M)$$

$$H_2 = H_r \cos(\theta_H - \theta_M) - 4\pi M_{eff} \cos(\theta_M)^2$$

The calculated FMR frequency $f_{FMR}$ dispersions of the FL and RL are illustrated in Fig.5 by red and black dashed lines, respectively. The FL $f_{FL-AP}$ and $f_{FL-P}$ modes have a lower frequency than or closing to their $f_{FMR}$ (red dashed line), while the RL $f_{RL-AP}$ mode has a higher frequency than its $f_{FMR}$ (black dashed line). Unlike the FMR frequency, the previous nonlinear auto-oscillator theory of microwave generation by spin-polarized current indicates that the STNO frequency is also affected by the nonlinear coefficient *N*, which highly depends on the PMA field and the angle and magnitude of the external magnetic field [42]. In addition, the studied MTJ is expected to have a significant voltage-controlled magnetic anisotropy coefficient, and the bias voltage corresponding to a large driving current will also change PMA and oscillating frequency[43].

## IV. CONCLUSIONS

In summary, our field- and current-driven magnetization switching and ST-FMR experiments reveal that the FL and RL have a strong interfacial PMA field ($H_\perp = 13.88\ kOe$ for the FL, $11.32\ kOe$ for the RL) and suffer a considerable stary magnetic field of 0.96 kOe from the bottom SyFi-PL layer. Besides the stary field, the RL also suffers an additional bias field of 1.16 kOe due to the moderate IEC between the RL and SyFi-PL. Furthermore, the FL and RL exhibit distinct dynamics with opposite *f* vs. *H* dispersion relations. The current-driven magnetization auto-oscillation for the FL can be achieved at both P and AP configurations but for the RL only at the AP state. Our results demonstrate a frequency crossover behavior for the FL and RL modes at the AP state when *H* is above 1.47 kOe at certain oblique angles. Our results provide valuable information for the selective excitation and electronic control of coherent magnetization auto-oscillation of a pMTJ-based STNO for *rf* applications and magnon-based neuromorphic computing by tunning the magnetization configurations between the FL and RL and their magnetic properties, e.g., various effective bias fields through tailoring the material parameters of the pMTJ multilayer structure.

### ACKNOWLEDGMENTS
We acknowledge support from the National Natural Science Foundation of China (Grants No. 12074178, No. 12004171, and No. 11874135), the Applied Basic Research

Programs of Science and Technology Commission Foundation of Jiangsu Province (Grant No. BK20200309), the Open Research Fund of Jiangsu Provincial Key Laboratory for Nanotechnology, the Scientific Foundation of Nanjing University of Posts and Telecommunications (NUPTSF) (Grant No. NY220164), and Key Research and Development Program of Zhejiang Province (Grant No. 2021C01039).